\documentclass[prl,nofootinbib,twocolumn,letterpaper,showpacs,showkeys]{revtex4}
\usepackage{amsmath,amssymb,times,graphics,graphicx,bm}

\newcommand{\be}{\begin{equation}}
\newcommand{\ee}{\end{equation}}
\newcommand{\ben}{\begin{eqnarray}}
\newcommand{\een}{\end{eqnarray}}
\newcommand{\bes}{\begin{subequations}}
\newcommand{\ees}{\end{subequations}}
\newcommand{\bF}{\begin{figure}}
\newcommand{\eF}{\end{figure}}
\newcommand{\dg}{\dagger}

\input{Qcircuit}

\def\tr{ {\rm{Tr }}}

\newcommand{\Proj}[1]{\mbox{$|#1\rangle \!\langle #1 |$}}

\begin{document}
\title{Quantum discord and the power of one qubit}
\author{Animesh Datta}
\email{animesh@unm.edu}
\affiliation{Department of Physics and Astronomy, MSC07--4220, 1
University of New Mexico, Albuquerque, New Mexico 87131-0001}
\author{Anil Shaji}
\email{shaji@unm.edu}
\affiliation{Department of Physics and Astronomy, MSC07--4220, 1
University of New Mexico, Albuquerque, New Mexico 87131-0001}
\author{Carlton M.~Caves}
\affiliation{Department of Physics and Astronomy, MSC07--4220, 1
University of New Mexico, Albuquerque, New Mexico 87131-0001}

\date{\today}
\begin{abstract}
We use quantum discord to characterize the correlations present in
the quantum computational model DQC1, introduced by Knill and
Laflamme [Phys.\ Rev.\ Lett.\ {\bfseries 81}, 5672 (1998)].  The
model involves a collection of qubits in the completely mixed state
coupled to a single control qubit that has nonzero purity.  The
initial state, operations, and measurements in the model all point to
a natural bipartite split between the control qubit and the mixed
ones.  Although there is no entanglement between these two parts, we
show that the quantum discord across this split is nonzero for
typical instances of the DQC1 ciruit.  Nonzero values of discord
indicate the presence of nonclassical correlations.  We propose
quantum discord as figure of merit for characterizing the resources
present in this computational model.
\end{abstract}


\maketitle

Characterizing and quantifying the information-processing
capabilities offered by quantum phenomena like entanglement,
superposition, and interference is one of the primary objectives of
quantum information theory.  In spite of substantial progress
\cite{ekert98a,jozsa03a,kendon06a}, the precise role of entanglement
in quantum information processing remains an open question
\cite{bennett99c,braunstein99a,vidal03b,biham04a,kenigsberg06a,datta07a}.
It is quite well established that entanglement is essential for
certain kinds of quantum-information tasks like quantum cryptography
and super-dense coding. In these cases, it is also known that the
quantum enhancement must come from entanglement spread over large
parts of the system.  It is not known, however, if all
information-processing tasks that can be done more efficiently with a
quantum system than with a comparable classical system require
entanglement as a resource.

For pure-state quantum computation, it is known that entanglement
must grow with the system size for there to be exponential speedup
\cite{jozsa03a}. There is evidence that quantum information
processors using highly mixed states, with no discernible
entanglement, can perform better
\cite{biham04a,braunstein99a,kenigsberg06a,meyer00a} than equivalent
classical ones.  Indeed, there exist models of mixed-state quantum
computation that provide exponential speedup over the best known
classical algorithms and yet have a bounded amount of entanglement
\cite{knill98a}.  Here we explore an alternate way of characterizing
the quantum nature of the correlations in such systems.

Quantum discord, introduced by Ollivier and Zurek and independently
by Henderson and Vedral~\cite{henderson01a,ollivier01a}, captures the
nonclassical correlations, including but not limited to entanglement,
that can exist between parts of a quantum system.  We investigate
the effectiveness of discord in characterizing the performance of the
model of quantum information processing introduced by Knill and
Laflamme in~\cite{knill98a}, which is often referred to as the {\em
power of one qubit}, or DQC1.  In this model, information processing
is performed with a collection of qubits in the completely mixed
state coupled to a single control qubit that has some nonzero purity.
Such a device can perform efficiently certain computational tasks for
which there is no known efficient method using classical information
processors.

We start with a discussion of quantum discord, its definition and its
relevance in quantum information theory. Consider the following two-qubit
separable state
\begin{eqnarray}
\label{eq:state1} \rho &=&\frac{1}{4}\Big(| + \rangle \langle +  |
\otimes |0 \rangle  \langle 0| \; + \; |- \rangle \langle -| \otimes |1 \rangle
\langle 1 |\nonumber \\
&& \hspace{0.5 cm} + |0\rangle \langle 0 | \otimes |- \rangle \langle -| \;  +
\; | 1 \rangle \langle 1|  \otimes |+ \rangle \langle  +| \Big),
\end{eqnarray}
in which four nonorthogonal states of the first qubit are correlated
with four nonorthogonal states of the second qubit.  Such
correlations cannot exist in any classical state of two bits.  The
extra correlations the quantum state can contain compared to an
equivalent classical system with two bits could reasonably be called
quantum correlations.  Entanglement is a special kind of quantum
correlation, but not the only kind.  In other words, separable quantum
states can have correlations that cannot be captured by a probability
distribution defined over the states of an equivalent classical
system.

Quantum discord attempts to quantify all quantum correlations
including entanglement.  It must be emphasized here that the discord,
in a sense, supplements the measures of entanglement that can be
defined on the system of interest.  It aims to capture all the
nonclassical correlations present in a system, those that can be
identified as entanglement and then some more.

The information-theoretic measure of correlations between two systems
$S$ and $M$ is the mutual information,
\begin{equation}
\label{eq:mutualI}
\mathcal{I}(S:M) = H(S) + H(M)- H(S,M).
\end{equation}
If $M$ and $S$ are classical systems whose state is described by a
probability distribution $p(S,M)$, then $H(\cdot)$ denotes the
Shannon entropy, $H(\vec p\,)\equiv -\sum_j p_j \log p_j$.  If $M$
and $S$ are quantum systems described by a combined density matrix
$\rho_{SM}$, then $H(\cdot)$ stands for the corresponding von Neumann
entropy, $H (\rho) \equiv -\tr (\rho \log \rho)$.

For classical probability distributions, Bayes's rule leads to an
equivalent expression for the mutual information,
\begin{equation}
\label{classicalJ}
\mathcal{I}(S:M) = H(S) - H(S|M),
\end{equation}
where the conditional entropy $H(S|M)$ is an average of Shannon
entropies for $S$, conditioned on the alternatives for $M$.  For
quantum systems, we can regard Eq.~(\ref{classicalJ}) as defining a
conditional entropy, but it is not an average of von Neumann
entropies and is not necessarily nonnegative~\cite{cerf99b}.

Another way of generalizing the classical conditional entropy to the
quantum case is to recognize that classically $H(S|M)$ quantifies the
ignorance about the system $S$ that remains if we make measurements
to determine $M$.  When $M$ is a quantum system, the amount of
information we can extract about it depends on the choice of
measurement.  If we restrict to projective measurements described by
a complete set of orthogonal projectors, $\{\Pi_j\}$, corresponding to
outcomes~$j$, then the state of $S$ after a measurement is given by
\begin{equation}
\rho_{S|j} = \frac{\tr_M\bigl(\Pi_j\rho_{SM}\Pi_j\bigr)}{p_j},
\qquad
p_j=\tr_{S,M}\bigl(\rho_{SM}\Pi_j\bigr).
\end{equation}
A quantum analogue of the conditional entropy can then be defined as
$\tilde{H}_{\{\Pi_j\}}(S|M) \equiv \sum_j p_j H(\rho_{S|j})\ge0$.  Since
$\rho_S=\sum_jp_j\rho_{S|j}$, the concavity of von Neumann entropy
implies that $H(S)\ge\tilde H_{\{\Pi_j\}}(S|M)$.  We can now define an
alternative quantum version of the mutual information,
\begin{equation}
\label{quantJ}
\mathcal{J}_{\{\Pi_j\}}(S:M) \equiv H(S) - \tilde H_{\{\Pi_j\}}(S|M)\ge 0.
\end{equation}
Performing projective measurements onto a complete set of orthogonal
states of $M$ effectively removes all nonclassical correlations
between $S$ and $M$.  In the post-measurement state, mutually
orthogonal states of $M$ are correlated with at most as many states
of $S$. It is easy to see that these sorts of correlations can be
present in an equivalent classical system.

The value of $\mathcal{J}_{\{\Pi_j\}}(S:M)$ in Eq.~(\ref{quantJ}) depends
on the choice of $\{\Pi_j\}$.  We want $\mathcal{J}_{\{\Pi_j\}}(S:M)$ to
quantify {\em all\/} the classical correlations in $\rho_{SM}$, so we
maximize $\mathcal{J}_{\{\Pi_j\}}(S:M)$ over all $\{\Pi_j\}$ and define a
measurement-independent mutual information
\begin{equation}
\label{quantJ1}
\mathcal{J}(S:M) \equiv H(S) - \tilde H(S|M)\ge0,
\end{equation}
where
\begin{equation}
\tilde H(S|M)\equiv \min_{\{\Pi_j\}}\sum_j p_j H(\rho_{S|j})
\end{equation}
is a measurement-independent conditional information.  The quantum
discord is then defined as
\begin{eqnarray}
\mathcal{D}(S,M) & \equiv & \mathcal{I}(S:M)-\mathcal{J}(S:M) \nonumber \\
&=& H(M)-H(S,M) + \tilde H(S|M) \nonumber \\
&=& \tilde H(S|M)-H(S|M).
\end{eqnarray}
The discord is nonnegative and is zero for states with only classical
correlations~\cite{henderson01a,ollivier01a}.  Thus a nonzero value
of $\mathcal{D}(S,M)$ indicates the presence of nonclassical
correlations~\cite{ollivier01a}.  The discord is bounded above by the
marginal entropy $H(M)$~\cite{proof}.

When the joint state $\rho_{SM}$ is pure, $H(S,M)$ and $\tilde
H(S|M)$ are zero, $H(S)=H(M)=-H(S|M)$, and the discord is equal to
$H(M)$, which is a measure of entanglement for bipartite pure states.
In other words, for pure states all nonclassical correlations
characterized by quantum discord can be identified as entanglement as
measured by the marginal entropy.

So far we have seen how discord can be used to characterize the
nonclassical nature of the correlations in quantum states.  We now
apply these ideas to the DQC1 or {\em power-of-one-qubit\/}
model~\cite{knill98a} of mixed-state quantum computation, which
accomplishes the task of evaluating the normalized trace of a unitary
matrix efficiently. The quantum circuit corresponding to this model
has a collection of $n$ qubits in the completely mixed state,
$I_n/2^n$, coupled to a single pure control qubit.  A generalized
version of this quantum circuit, with the control qubit having
sub-unity polarization, is shown below: \vspace{-9pt}
\begin{equation*}
 \Qcircuit @C=.5em @R=-.5em {
    & \lstick{{\frac{1}{2}}(I_1+\alpha Z)}
        & \gate{H} & \ctrl{1} & \meter & \push{\rule{0em}{4em}} \\
    & & \qw & \multigate{4}{U_n} & \qw & \qw \\
    & & \qw & \ghost{U_n} & \qw & \qw \\
    & \lstick{\mbox{$I_n/2^n\Big\{$}} & \qw & \ghost{U_n} & \qw & \qw \\
    & & \qw & \ghost{U_n} & \qw & \qw \\
    & & \qw & \ghost{U_n} & \qw & \qw
}
\end{equation*}
This circuit evaluates the normalized trace of $U_n$,
$\tau=\tr(U_n)/2^n$, with a polynomial overhead going as
$1/\alpha^2$.

The problem of evaluating $\tau$ is believed to be hard classically.
Quantum mechanically, the circuit provides an estimate of $\tau$ up
to a constant accuracy in a number of trials that does not scale
exponentially with $n$.  It does so by making $X$ and $Y$
measurements on the top qubit.  The averages of the obtained binary
values provide estimates for $\tau_R \equiv \mathrm{Re}(\tau)$ and
$\tau_I \equiv \mathrm{Im}(\tau)$.  The top qubit is completely
separable from the bottom mixed qubits at all times.  The final state
has vanishingly small entanglement, as measured by the
negativity~\cite{datta05a} across any split that groups the top qubit
with some of the mixed qubits.  Nonetheless, there is evidence that
the quantum computation performed by this model cannot be simulated
efficiently using classical computation~\cite{datta07a}.

The DQC1 circuit transforms the highly-mixed initial state $\rho_0
\equiv \Proj{0}\otimes I_n/2^n$ into the final state $\rho_{n+1}$,
\begin{eqnarray}
 \rho_{n+ 1} &=& \frac{1}{2^{n+1}} \Bigl( |0\rangle\langle0| \otimes I_n +
|1\rangle\langle1| \otimes I_n  \nonumber\\
 & & \hspace{1 cm} + \; \alpha |0\rangle \langle1| \otimes U_n^\dagger + \alpha
|1\rangle\langle0| \otimes U_n \Bigr)\nonumber\\
 &=&\frac{1}{2^{n+1}}\left(%
\begin{array}{cc}
  I_n & \alpha U^{\dg}_n \\
  \alpha U_n & I_n \\
\end{array}%
\right). \label{eq:rhoT}
\end{eqnarray}

Everything about the DQC1 setup, including the measurements on the
control qubit, suggests a bipartite split between the control qubit
$M$ and the mixed qubits $S$.  Relative to this split, we turn to
computing the quantum discord for the state $\rho_{SM}=\rho_{n+1}$.
The joint state $\rho_{n+1}$ has eigenvalue spectrum
\begin{equation}
\label{fullspectrum}
\vec{\lambda}(\rho_{n+1})=\frac{1}{2^{n+1}}(\underbrace{1-\alpha,\cdots,1-\alpha
} _ { 2^n
\mathrm{times}},\underbrace{1+\alpha,\cdots,1+\alpha}_{2^n
\mathrm{times}}),
\end{equation}
which gives a joint entropy
\begin{equation}
H(S,M)=n+H_2\bigg(\frac{1-\alpha}{2}\bigg).
\end{equation}
The marginal density matrix for the control qubit at the end of the
computation is
 \begin{equation}
 \rho_M = \frac{1}{2}\left(%
\begin{array}{cc}
  1 & \alpha\;\tau^* \\
  \alpha \;\tau & 1 \\
\end{array}%
\right), \label{eq:rhoM}
 \end{equation}
which has eigenvalues $(1\pm \alpha|\tau|)/2$ and entropy
 \begin{equation}
H(M) = H_2\bigg(\frac{1-\alpha|\tau|}{2}\bigg),
 \end{equation}
where $H_2(\cdot)$ is the binary Shannon entropy.

The evaluation of the quantum conditional entropy involves a
minimization over all possible one-qubit projective measurements. The
projectors are given by $ \Pi_\pm =
\frac{1}{2}(I_1\pm\bm{a\cdot\sigma})$, with $\bm{a\cdot
a}=a_1^2+a_2^2 + a_3^2 =1.$  The post-measurement states are
 \begin{equation}
 \rho_{S|\pm}=\frac{1}{p_\pm 2^{n+1}}\bigg(
I_n \pm \alpha\frac{a_1-i a_2}{2} U_n \pm \alpha\frac{a_1 + i a_2}{2}
U_n^\dg \bigg) ,
 \end{equation}
occurring with outcome probabilities
\begin{equation}
 p_\pm = \frac{1}{2}[1\pm\alpha(a_1 \tau_R + a_2 \tau_I)].
\end{equation}
The post-measurement states are independent of $a_3$, so without loss
of generality, we can let $a_3=0$, $a_1= \cos\phi$, and
$a_2=\sin\phi$.  The corresponding post-measurement states are
\begin{equation}
\rho_{S|\pm} = \frac{1}{p_\pm 2^{n+1}} \bigg(
I_n \pm \alpha\frac{e^{-i\phi} U_n + e^{i \phi} U_n^{\dagger}}{2}
\bigg) ,
\end{equation}

To find the discord of the state at the end of the computation, we
need the spectrum of $\rho_{S|\pm}$ so that we can compute
$H(\rho_{S|\pm})$.  The eigenvalues of any unitary operator $U_n$ are
phases of the form $e^{i\theta_k}$, so we have
\begin{equation}
\lambda_k \bigg( \frac{e^{-i\phi} U_n + e^{i\phi} U_n^{\dg}}{2} \bigg)
=\cos(\theta_k-\phi), \quad k=1,\cdots, 2^n,
\end{equation}
and
\begin{equation}
\label{postspectrum}
\lambda_k(\rho_{S|\pm})=
\frac{1}{2^n}
\frac{1\pm\alpha\cos(\theta_k-\phi)}{1\pm\alpha(\tau_R\cos\phi+\tau_I\sin\phi)}
\equiv q_{k\pm}.
\end{equation}
We also have
\begin{equation}
\tau_R = \frac{1}{2^n} \sum_k \cos \theta_k
\quad\mbox{and}\quad
\tau_I = \frac{1}{2^n} \sum_k \sin\theta_k.
\end{equation}
All this gives $H(\rho_{S|\pm})=H(\vec q_\pm)$ and thus
\begin{eqnarray}
\label{condent}
\tilde H_{\Pi_\pm}& =& p_+ H(\rho_{S|+})+p_- H(\rho_{S|-}\nonumber \\
&=&\frac{1}{2}[H(\vec{q}_+)+H(\vec{q}_-)]\nonumber\\
&&+\frac{\alpha}{2}(\tau_R\cos\phi +
\tau_I\sin\phi)[H(\vec{q}_+)-H(\vec{q}_-)].\qquad
\end{eqnarray}

We now use the fact that we are interested in the behavior of the
quantum discord of the DQC1 state for a typical unitary. By typical,
we mean a unitary chosen randomly according to the (left and right
invariant) Haar measure on $\mathbb{U}(2^n)$.  For such a unitary, it
is known that the phases $\theta_k$ are almost uniformly distributed
on the unit circle with large probability \cite{diaconis03a}.  Thus
for typical unitaries $\sum_k e^{i\theta_k}$ is close to zero.  Hence
both $\tau_R$ and $\tau_I$ are small, and we can ignore the second
term on the right-hand side in Eq.~(\ref{condent}).  In addition, the
phases $\theta_k$ can be taken to be placed at (with large
probability) the $2^n$-th roots of unity, i.e., $\theta_k = 2\pi
k/2^n$.  It follows that the spectra $\lambda_k(\rho_{S|\pm})$ are
independent of $\phi$.  Hence the entropies we are interested in
computing are also independent of $\phi$, and we can set $\phi$ to
zero without loss of generality.  This choice for $\phi$ corresponds
to measuring the pure qubit $M$ along $X$. The $X$ measurement gives
the real part of the normalized trace of $U_n$, and it is one of the
two measurements discussed in the original proposal by Knill and
Laflamme. Setting $\phi=\pi/2$ yields the other measurement, along
$Y$, which gives the imaginary part of the normalized trace of $U_n$.

In the limit of large $n$, we can simplify Eq.~(\ref{condent}) as
follows:
\begin{eqnarray}
\label{eq:sum}
\tilde H&=&  \frac{1}{2}[H(\vec{q}_+)+H(\vec{q}_-)] \nonumber \\
&=& -\frac{1}{2^{n+1}}\sum_{k=1}^{2^n}\bigg[(1+\alpha
\cos\theta_k) \log\bigg(\frac{1+\alpha \cos\theta_k}{2^n}\bigg) \nonumber \\
&& \hspace{1 cm} + \; (1-\alpha \cos\theta_k)\log \bigg( \frac{1-\alpha
\cos \theta_k}{2^n} \bigg) \bigg]\nonumber \\
&=& n - \frac{1}{2^{n+1}} \sum_{k=1}^{2^n} \bigg[ \log \big(1-\alpha^2
\cos^2\theta_k \big) \nonumber\\
&& \hspace{1 cm} + \;\alpha \cos\theta_k\log \bigg( \frac{1+\alpha
\cos\theta_k}{1-\alpha \cos\theta_k}\bigg) \bigg].
\end{eqnarray}
Furthermore, when $n$ is large, we can replace the sum in the above
equation with an integral to obtain
\begin{eqnarray}
\tilde H&=&n-\frac{1}{4\pi}\bigg[\int_0^{2\pi}\log(1-\alpha^2 \cos^2x)
\mathrm{d}x \nonumber \\
&& \hspace{1 cm} + \; \alpha \int_0^{2\pi} \cos x\log \bigg( \frac{1+\alpha \cos
x}{1-\alpha \cos x} \bigg) \mathrm{d}x \bigg] \nonumber \\
&=& n +1 - \log\Big(1+\sqrt{1-\alpha^2}\Big) \nonumber \\
&& \hspace{2.2 cm} -\Big(1-\sqrt{1-\alpha^2}\Big)\log e.
\end{eqnarray}
Note that when the sums are replaced by integrals,
$H(\vec{q}_+)-H(\vec{q}_-)=0$, providing further justification for
ignoring the second term in Eq.~(\ref{condent}).

\begin{figure}[!ht]
\begin{center}
\resizebox{8.5 cm}{5.8 cm}{\includegraphics{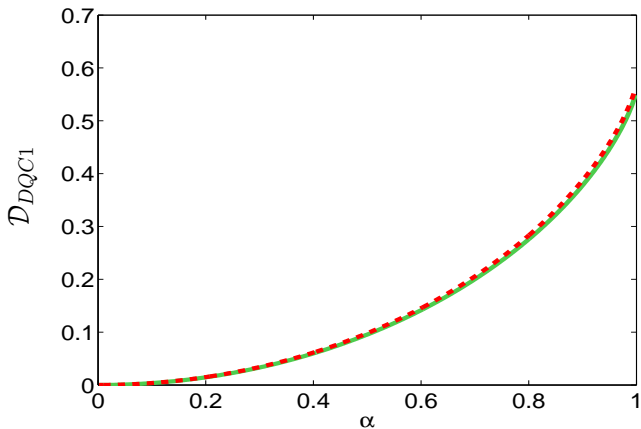}}
 \caption{(Color online) The dashed (red) line shows the average
discord in a DQC1 circuit with five qubits in the mixed state ($n=5$)
coupled to a qubit with purity $\alpha$.  The average is taken over
five hundred instances of pseudo-random unitary matrices.  The discord
is shown as a function of the purity of the control qubit.  The solid
(green) line shows the analytical expression in
Eq.~(\ref{randomdiscord}), which grows monotonically from 0 at
$\alpha=0$ (completely mixed control qubit) to $2-\log e=0.5573$ at
$\alpha=1$ (pure control qubit).  Even for $n=5$ the analytical
expression is quite accurate.  These values of discord should be
compared with a maximum possible discord of 1 when $M$ is a single
qubit.} \label{discorddqc}
\end{center}
\end{figure}

When $|\tau|$ is small, $H(M)\simeq 1$, and the quantum discord for
the DQC1 state is then given by the simple expression
\begin{eqnarray}
\label{randomdiscord}
\mathcal{D}_{\rm DQC1} &=& 2 - H_2 \Big(\frac{1-\alpha}{2} \Big)
- \log\Big(1+\sqrt{ 1-\alpha^2}\Big)\nonumber \\
&& \hspace{1 cm}-\Big(1-\sqrt{1-\alpha^2}\Big)\log e.
\end{eqnarray}
Figure~\ref{discorddqc} compares the discord from
Eq.~(\ref{randomdiscord}) with the average discord in a DQC1 circuit
having five qubits in the mixed state ($n=5)$ coupled to a control
qubit with purity $\alpha$.  The average is taken over 500 instances
of pseudo-random unitary matrices.  We see that in spite of the
approximations made in obtaining Eq.~(\ref{randomdiscord}), the
analytic expression provides a very good estimate of the discord even
when $n$ is as low as five.

There is no entanglement between the control qubit and the mixed
qubits in the DQC1 circuit at any point in the computation, yet there
are nonclassical correlations, as measured by the discord, between
the two parts at the end of the computation for any $\alpha > 0$.
Other bipartite splittings of $\rho_{n+1}$ can exhibit entanglement,
but it was shown in~\cite{datta05a} that the partial transpose
criterion failed to detect entanglement in $\rho_{n+1}$ for $\alpha
\leq 1/2$. In this domain, several other tests for entanglement,
including the first level of the scheme of
Doherty~\textit{et~al.}~\cite{doherty04a}, which is based on
semi-definite programming, also failed to detect entanglement. The
above expression is thus the first signature of nonclassical
correlations in the DQC1 circuit for $\alpha \leq 1/2$.

In conclusion, we calculated the discord in the DQC1 circuit and
showed that nonclassical correlations are present in the state at the
end of the computation even if there is no detectable entanglement.
This shows that for some purposes quantum discord might be a better
figure of merit for characterizing the quantum resources available to
a quantum information processor.  We present evidence of the presence
of nonclassical correlations in the DQC1 circuit when $\alpha \leq
1/2$. The quantum discord for qubits is known to be a true measure of
nonclassical correlations~\cite{hamieh04a}.  This suggests that
nonclassical correlations other than entanglement, as quantified by
the discord, may explain the (sometimes exponential) speed-up in the
DQC1 circuit and perhaps the speedup in other quantum computational
circuits also. For pure states, discord becomes a measure of
entanglement.  Therefore, using discord to connect quantum resources
to the advantages offered by quantum information processors has the
additional advantage that it works well for both pure- and
mixed-state quantum computation.

We thank W.~H. Zurek, C.~Rodriguez, and K.~Modi for useful
discussions on quantum discord.  This work was supported in part by
Office of Naval Research Contract No.~N00014-07-1-0304 and National
Science Foundation Grant No.~PHY-0653596.

\bibliography{discordDQC}

\begin{thebibliography}{18}
\bibitem{proof} The eigendecomposition $\rho_{SM}=\sum_a p_a\Pi_a$ 
yields $p_j\rho_{S|j}= \sum_a p_a p_{j|a}\rho_{S|a,j}$, where 
$\rho_{S|a,j}=\tr_M(\Pi_j\Pi_a)/p_{j|a}$ is a pure state of $S$.  It 
follows from the pure-state decomposition $\rho_{S|j}=\sum_a 
p_{a|j}\rho_{S|a,j}$ that $H(A|j)\ge S(\rho_{S|j})$.  Thus 
$H(S,M)=H(A)\ge H(A|J)=\sum_j p_j H(A|j)\ge\tilde 
H_{\{\Pi_j\}}(S|M)\ge\tilde H(S|M)$, from which the upper bound on 
discord follows.
\end{thebibliography}

\end{document}